\documentclass[aps,showpacs,preprintnumbers,amsmath,amssymb,superscriptaddress,floatfix,a4paper,nofootinbib,onecolumn]{revtex4}
\usepackage{graphicx}
\usepackage{bm}
\usepackage{rotating}
\usepackage{array}
\usepackage{xcolor}
\usepackage{amsmath,amssymb}
\usepackage{mathrsfs}
\usepackage{graphicx}
\usepackage{color}
\usepackage{orcidlink}
\usepackage{subfigure}
\usepackage{fancyhdr}
\usepackage{multirow}
\usepackage{float}
\usepackage{epsfig}
\usepackage{amsfonts}
\usepackage{bm}
\usepackage{multirow}
\usepackage{nicefrac}
\usepackage{cancel}
\newcommand{\ba}{\begin{eqnarray}}
\newcommand{\ea}{\end{eqnarray}}

\usepackage{bbm}
\newcommand{\be}{\begin{eqnarray}}
\newcommand{\ee}{\end{eqnarray}}
\usepackage{slashed}
\usepackage{booktabs} 

\begin{document}

\title{Analytical Solutions to the Wheeler–DeWitt Equation in Rosen-Lagrangian Cosmology via the Eisenhart Lift}

\author{Narakorn Kaewkhao\orcidlink{000-0002-1316-7200}}
\email{naragorn.k@psu.ac.th}
\affiliation{Division of Physical Science, Faculty of Science, Prince of Songkla University, Hatyai 90112, Thailand} 

\author{Suparat Marit}
\email{maritsuparat23@gmail.com}
\affiliation{Division of Physical Science, Faculty of Science, Prince of Songkla University, Hatyai 90112, Thailand} 

\author{Phongpichit Channuie\orcidlink{0000-0003-4489-1207}} 
\email{phongpichit.ch@mail.wu.ac.th}
\affiliation{College of Graduate Studies, Walailak University, Thasala, Nakhon Si Thammarat, 80160, Thailand}
\affiliation{School of Science, Walailak University, Thasala, Nakhon Si Thammarat, 80160, Thailand}

\date{\today}

\begin{abstract}
The Rosen Lagrangian framework promotes the cosmological constant to a scale-factor-dependent quantity, $\Lambda(a)=\Lambda_{0}a^{\lambda}$, thereby providing a dynamical dark energy scenario for $\lambda \neq 0$. In the special case $\lambda=0$, the model naturally reduces to the standard $\Lambda$CDM cosmology. Within this framework, the conformal Killing equations are employed to determine the conformal factor $\mathcal{F}(a)$, which is expressed in terms of the effective potential $V_{\rm eff}$ and its derivative $V'_{\rm eff}$. Furthermore, the Eisenhart lift formalism introduces an additional field $\chi$, allowing the cosmological dynamics to be reformulated through a purely kinetic lifted action. This geometrical construction provides a powerful approach to quantum cosmology by transforming the Wheeler–DeWitt equation into a tractable form that admits analytic solutions. Such solutions are particularly relevant in cosmological epochs dominated by the cosmological constant, including both the inflationary era of the early Universe and the late-time accelerated expansion. Consequently, this framework offers a promising avenue for connecting geometrical methods, quantum cosmology, and dynamical dark energy within a unified description.
\end{abstract}

\maketitle


\section{Introduction}
The concept of the Eisenhart lift process, which originates from Eisenhart’s work in mechanics \cite {Eisenhart origin}, can be applied to the cosmological Lagrangian proposed by Nathan Rosen, a renowned physicist, \cite{Rosen1993, Newtonian cosmo, Adiabatic invariant}, to explore deeper aspects of gravitational dynamics. Rosen made significant contributions to cosmology, including the formulation of a cosmological Lagrangian that captures the Universe dynamics within a Newtonian framework. The essential Rosen's cosmological Lagrangian\cite{Rosen1993}, which allows us to derive the Friedmann equation and the acceleration equation using Newtonian physics\cite{Newtonian cosmo}. We utilize the Noether symmetry method alongside the often-overlooked adiabatic invariant in cosmological models, as cited in Ref \cite{Adiabatic invariant}. In previous work, we demonstrated that the cyclic Universe exhibits an oscillation period of approximately $15.8 Gy$. In this work, we employ the Eisenhart–Duval lift method to reformulate the cosmological dynamics within a geometric framework. Using this approach, it becomes possible to determine the oscillation period of the Universe and to investigate the conditions under which cyclic or oscillatory cosmological behavior may emerge on a characteristic timescale comparable to the Hubble timescale. 
The reference \cite{Cariglia 2018} investigates the Eisenhart lift formalism for a single scalar field in an FLRW Universe using the Eternal Peripatetic form while retaining $\rho(t),k,$ and $\Lambda$. Furthermore, the Ermakov–Milne–Pinney equation can map the dynamics of a scalar field onto a simple harmonic oscillator with a time-dependent frequency. Ref. \cite{Georgios 2023} employs the Eisenhart lift framework to construct the mean curvature and uses the resulting geometric structure to characterize observable parameters in inflationary cosmology. In this work, we apply the Eisenhart lift formalism to the Rosen cosmological Lagrangian in the framework of Newtonian cosmology and investigate the new insights it provides. By introducing an additional variable, $\chi(t) $, the lifted Lagrangian acquires a purely kinetic form, which significantly simplifies the analysis of the Killing vectors. A distinctive feature of this construction is that $\chi(t)$ does not explicitly appear in the field-space metric $G_{AB},$ thereby acting as a cyclic variable in the Eisenhart-lift framework. Consequently, the conformal Killing equations become considerably more tractable, facilitating the determination of hidden symmetries associated with the Eisenhart-lifted Rosen Lagrangian.
\ba\label{Full Nathan Lagrangian}
L_{\rm Rosen}(a,\dot{a})&=&T-V,\\
&=&\frac{1}{2}m\dot{a}^{2}+\frac{4\pi G_{\rm N}}{3}ma^{2}\rho(a)+\frac{1}{6}ma^{2}\Lambda(a)  -\frac{mk }{2},\label{Full Nathan Lagrangian}
\ea
$\rho(a)$ denotes the energy density of matter and radiation only, while dark energy is treated separately through a time-dependent cosmological constant. The corresponding dark-energy density is defined by $ \rho_{\Lambda}=\frac{\Lambda(a)}{8\pi G_{\rm N}}$ where a power-law form of the cosmological constant, $\Lambda(a)=\Lambda_{0}a^{\lambda},$ has been assumed. Several interesting dynamical forms of the cosmological constant have been proposed in the literature (see, e.g., \cite{Macedo 2023,Koussour 2024,Overduin 1998,Carneiro 2021,Saibal Ray,Wei Chen 1990,Myrzakulov 2024}, and references therein), leading to a variety of significant cosmological implications beyond the standard  $\Lambda \rm CDM$ scenario. Motivated by these earlier developments, the present parametrization provides a systematic framework for exploring possible deviations from the model and for investigating dynamical dark-energy effects via an ansatz for $\Lambda(a)$. Using Eq.(\ref{Full Nathan Lagrangian}), we can write the related effective potential as
\ba\label{effective potential}
V_{\rm eff}(a)=
-\frac{4\pi G_{\rm N}}{3}ma^{2}\rho(a)-\frac{1}{6}ma^{2}\Lambda(a) +\frac{mk }{2}.
\ea
The Lagrangian in Eq.(\ref{Full Nathan Lagrangian}) incorporates gravitational kinetic and potential terms such that the resulting equations reproduce the Friedmann equations describing the expanding Universe.  The Rosen Lagrangian yields the Hamiltonian
\ba\label{Full Nathan Hamiltonian}
\mathcal{H}=0&=&T(a,\dot{a})+V_{\rm eff}(a), \nonumber\\
&=&\frac{p_{a}^{2}}{2m}+V_{\rm eff}(a).
\ea
where $p_{a}=\frac{1}{2}m\dot{a}^{2}.$ By dividing Eq.(\ref{Full Nathan Hamiltonian}) by $\frac{m\dot{a}^{2}}{2}$ and applying the constraint $\mathcal{H}=0,$ the standard Friedmann equation \cite{A Liddle 2015} is recovered as follows:
\ba\label{Hubble constant}
H^{2}=\frac{8\pi G_{\rm N}}{3}\rho(a)+\frac{\Lambda(a) }{3}-\frac{k}{a^{2}},
\ea
where the Hubble parameter is defined as $H=\frac{\dot{a}}{a}.$
Taking the partial derivative $V_{\rm eff}$ with respect to the scale factor a(t), this gives
\ba
V_{\rm eff,a}=\frac{\partial V_{\rm eff}}{\partial a}&=& -\frac{8\pi G m a \rho }{3}+\frac{4}{3}\pi G_{\rm N}ma3(\rho+p)-\frac{ma\Lambda_{0}a^{\lambda}}{3}-\frac{1}{6}ma^{2}\Lambda_{0}\lambda a^{\lambda-1}, \nonumber \\
&=& \frac{4\pi G_{\rm N}ma\rho}{3}(1+3w)-\frac{ma\Lambda(a)}{3}-\frac{1}{6}m\lambda a\Lambda(a),\label{dif V-eff2}
\ea
Here, the fluid equation governing the evolution of the matter field (valid for radiation, ordinary matter, and dark matter), 
$\frac{\partial \rho}{\partial a}=-\frac{3}{a}(\rho+p)$, along with the equation of state $p(a)=w\rho(a)$
, which $w=\frac{p(a)}{\rho(a)}$ is assumed to be constant at certain particular times, which are determined by the observational data. 
Including the energy density associated with the dynamical cosmological constant leads to an effective equation-of-state parameter that evolves gradually over cosmological timescales, given by
$w_{\rm eff }(t)=\frac{p_{R}+w_{\Lambda}\rho_{\Lambda}}{\rho_{m}+\rho_{R}+\rho_{\Lambda}}$ where $w_{\Lambda}(t)=\frac{p_{\Lambda}}{\rho_{\Lambda}}$ is the time-dependent equation-of-state parameter of dark energy. Using Newton’s equation of motion together with the Rosen cosmological force defined by the gradient of the effective potential~\cite{Rosen1993,Newtonian cosmo}, we obtain
\ba\label{the cosmological force}
F(a)=m\ddot{a}=-\frac{\partial V_{\rm eff}(a)}{\partial a}.
\ea
We once got the acceleration equation as follows
\ba\label{accel eq1}
\frac{\ddot{a}}{a}=-\frac{4\pi G_{\rm N}\rho}{3}(1+3w)+\frac{\Lambda_{0}}{3}a^{\lambda}+\frac{1}{6}\Lambda_{0}\lambda a^{\lambda}.
\ea 
By setting $\lambda=0$, this gives the standard $\Lambda \rm CDM$  model
\ba\label{accel eq std}
\frac{\ddot{a}}{a}=-\frac{4\pi G_{\rm N}\rho}{3}(1+3w)+\frac{\Lambda_{0}}{3}.
\ea 
The results presented in this section highlight the remarkable structure of the Rosen Lagrangian, which reproduces the fundamental equations of standard cosmology and yields solutions for key dynamical quantities, such as the scale factor. Motivated by this, we investigate the consequences of applying the Eisenhart lift framework to the Rosen Lagrangian with the ansatz $\Lambda(a)=\Lambda_{0}a^{\lambda}.$ In particular, we examine the role of the auxiliary field $\chi$ in extending the Eisenhart metric $G_{AB}$ to two dimensions, formulate the corresponding Wheeler–DeWitt equation within the Eisenhart lift approach, and explore possible analytical solutions.

The structure of this paper is as follows. In Section \ref{ch3}, we provide a brief overview of the Eisenhart lift formalism and apply it to the Rosen cosmological Lagrangian, thereby reformulating the original dynamical system in terms of a purely kinetic lifted action. In Section \ref{ch4}, we investigate the geometrical properties of the resulting minisuperspace, derive the geodesic equations, and analyze the associated conformal Killing equations, leading to the determination of the conformal factor and conserved quantities of the system. In Section \ref{ch5}, we formulate the Wheeler-DeWitt equation within the Eisenhart-lift framework and obtain analytical quantum cosmological solutions in terms of Bessel functions. Finally, in Section \ref{ch6}, we summarize our main results and discuss their implications for quantum cosmology, cyclic cosmological evolution, and possible future investigations of the Eisenhart lift approach in gravitational and cosmological theories.

\section{Eisenhart Lift: Concept and Overview }\label{ch3}
Eisenhart’s lifting technique, originally developed in the 1930s, provides a geometric framework that maps a Newtonian dynamical system onto a higher-dimensional manifold whose trajectories are geodesics. In this approach, the phase space of the original Newtonian system is extended to a higher-dimensional space, enabling the dynamics to be analyzed geometrically \cite {Finn 2023 Book,Cariglia 2015,K Fin 2018,Finn 2019}.
At this stage, we have all the necessary ingredients to reformulate the original Lagrangian into a purely kinetic form by applying the Eisenhart lift within the finite-dimensional minisuperspace configuration space, as shown below
\ba\label{new Lagrangian}
L_{\rm Lift}(a,\dot{a},\dot{\chi})&= \frac{1}{2}m\dot{a}^2 + \frac{1}{2} \frac{M^2\dot{\chi}^2}{V_{\rm eff}(a)} \\[1em]
&= \frac{1}{2} G_{AB} \dot{\Phi}^A \dot{\Phi}^B,
\ea
where $\Phi = \{a,\chi\}$ and $\dot{\Phi} = \{\dot{a},\dot{\chi}\}.$ 
In this framework, $a(t)$ and $\chi(t)$  are treated as independent configuration-space variables; therefore, derivatives such as $\frac{da}{d\chi}$ (and equivalently $\frac{d\chi}{da})$ vanish.  

The absence of $\chi$ from the Lagrangian indicates that $\chi$  is a cyclic variable, implying the conservation of the conjugate canonical momentum. An important feature of the Eisenhart lift is that the introduction of additional dynamical variables depends sensitively on the structure of the effective potential, as shown in Eq.(\ref{effective potential}). In the present treatment, the lift metric and its corresponding inverse are given by:
\begin{equation}
\begin{array}{c c}
G_{AB} = 
\begin{bmatrix}
m & 0 \\
0 & \frac{M^2}{V_{\text{eff}}(a)}
\end{bmatrix}
&,
G^{AB} = 
\begin{bmatrix}
\frac{1}{m} & 0 \\
0 & \frac{V_{\text{eff}}(a)}{M^2}
\end{bmatrix}
\end{array},
\end{equation}
Here, $\frac{M^{2}}{V_{\rm eff}}$ has the dimension of mass. In natural units  $c=1$, $M$ itself represents a mass scale. The new kinetic term obtaines from the original potential term by introducing the time derivative of the new dynamical variable, $\dot\chi(t)$. 
We apply the Euler-Lagrange equation to the scale factor $a$ :
\ba\label{EL a}
\frac{d}{dt}\Big(\frac{\partial L_{\rm Lift}}{\partial \dot{a}}\Big)-\frac{\partial L_{\rm Lift}}{\partial a}=0.
\ea
This gives
\ba\label{EL a2}
\ddot{a}=-\frac{1}{2}\frac{M^{2}}{m}\frac{\dot{\chi}^{2}V_{\rm eff,a}}{V_{\rm eff}^{2}}.
\ea
Likewise, we use the Euler-Lagrange equation to the extra field  $\chi$: 
\ba\label{EL for phi}
\frac{d}{dt}\Big(\frac{\partial L_{\rm Lift}}{\partial \dot{\chi}}\Big)-\cancel{\frac{\partial L_{\rm Lift}}{\partial \chi}}=0.
\ea
It implies that
\ba
\frac{d p_{\chi}}{dt}=0, \quad p_{\chi}\equiv \frac{\partial L_{\rm Lift}}{\partial \dot{\chi}}.
\ea
From this, one obtains
\ba\label{EL phi}
\frac{d}{dt}\Big(\frac{M^{2}\dot{\chi}}{V_{\rm eff}(a)}\Big)&=&0,\\
p_{\chi}=\frac{M^{2}\dot{\chi}}{V_{\rm eff}(a)}&=& AM = \rm const.  \label{phi eq}
\ea
To satisfy Eq.(\ref{accel eq1}), we must impose the Eisenhart condition 
$A=\sqrt{2}$ \cite{Finn 2023 Book}.  
By defining the useful quantity:
\ba
\dot{\chi}=\frac{\sqrt{2}V_{\rm eff}(a)}{M}. \label{link Lagrangian}
\ea
Eq. (\ref{link Lagrangian}) allows us to transform the Lift Lagrangian back into the original Rosen form, from which 
by inserting $V_{\rm eff},\,V_{\rm eff,a}, A=\sqrt{2},$  and the matter density evolution $\rho(a)=\rho_{0}a^{-3(1+w)}$ into Eq.(\ref{EL a2}), we once again obtain the acceleration equation.
\section{Geodesic Equations and Conformal Killing Equations}\label{ch4}
Next, we analyze the geodesic equation using the lift Lagrangian. The lifted Hamiltonian maps the cosmological dynamics to geodesic motion on a two-dimensional minisuperspace:
\ba
ds^{2}=m da^{2}+\frac{M^{2}}{V_{\rm eff}}d\chi^{2}.
\ea
The non-vanishing Christoffel symbols  derive from 
\ba
\Gamma^{C}_{AB}=\frac{1}{2}G^{CD}\Big( \partial_{A} G_{BD}+\partial_{B}G_{AD}-\partial_{D}G_{AB}  \Big).
\ea
The obtained result is as follows:
\ba\label{CTF}
\Gamma^{a}_{\chi\chi}&=&\frac{M^{2}}{2m}\frac{V_{\rm eff,a}}{V^{2}_{\rm eff}},   \\
\Gamma^\chi_{\chi a }=\Gamma^\chi_{a\chi }&=&-\frac{V_{\rm eff,a}}{2V_{\rm eff}}.
\ea
The generalized geodesic equation on the field space manifold shows
\ba
\ddot{\Phi}^{A}+\Gamma^{A}_{BC}\dot{\Phi}^{B}\dot{\Phi}^{C}=-G^{AB}\partial_{B}\mathcal{W}. \label{geodesic eq lift}
\ea
Assuming $\mathcal{W}(a,\chi)$, the term $-G^{AB}\partial_{B}\mathcal{W}$ on the right-hand side of (\ref{geodesic eq lift}) can be incorporated. Thus, the geodesic equation for the variable a takes the form: 
\ba
\ddot{a} + \Gamma^a_{\chi\chi} \dot{\chi}^{2}+\cancel{\Gamma^a_{\chi a}} \dot{\chi}\dot{a}+\cancel{\Gamma^a_{a\chi}}\dot{a} \dot{\chi} = -G^{aa}{\partial_{a}\mathcal{W}}- \cancel{G^{a\chi} \partial_\chi \mathcal{W}} &= &0, \label{ geodesic for a } \nonumber\\
\ddot{a}+\frac{M^{2}V_{\rm eff,a} \dot{\chi}^{2} }{2m V^{2}_{\rm eff}} + \frac{1}{m} {\partial_{a}\mathcal{W}(a,\chi)} &=& 0, \label{geodesic for a2} \nonumber\\
\ddot{a}+\frac{V_{\rm eff,a}}{m}+\frac{1}{m}\partial_{a}\mathcal{W}(a,\chi)&=&0.\label{Killing a}
\ea
Dividing both sides of Eq.(\ref{Killing a}) by $a$ and assuming $\mathcal{W}= \rm {const}$, this yields once again the acceleration equation of the Universe as shown in Eq.(\ref{accel eq1} ). 
Meanwhile, the second geodesic equation for variable $\chi$ takes the form:
\ba
\ddot{\chi} - \frac{V_{\rm eff,a}}{V_{\rm eff}} \dot{a} \dot{\chi} = -G^{\chi\chi}{\partial_{\chi}\mathcal{W}} &=& 0, \nonumber\\
\ddot{\chi} -\frac{A\dot{a}V_{\rm eff,a}}{M}+\frac{V_{\rm eff}(a)}{M^{2}}\partial_{\chi}\mathcal{W}(a,\chi)&=&0.\label{geodesic for chi}\\
\cancel{\frac{A\dot{a}V_{\rm eff,a}}{M}-\frac{A\dot{a}V_{\rm eff,a}}{M}}+\frac{V_{\rm eff}(a)}{M^{2}}\partial_{\chi}\mathcal{W}(a,\chi)&=&0,\label{no information}\\
\partial_{\chi}\mathcal{W}(a,\chi)=0, \quad \Rightarrow \mathcal{W}=\mathcal{W}(a).
\nonumber
\ea
Under the condition $\mathcal{W}=\rm {const}$ from Eq.(\ref{Killing a}),
(required by Eq. (\ref{Killing a})), Eq.(\ref{no information}) provides no further constraints at this stage. We now consider the lifted Hamiltonian associated with the lifted Lagrangian $L(a,\dot{a},\dot{\phi})$, given by
\ba
\mathcal{H}_{\rm Lift}&=&\frac{1}{2}G^{AB}p_{A}p_{B},\\ &=&\frac{1}{2}G^{aa}p^{2}_{a}+\frac{1}{2}G^{\chi\chi}p^{2}_{\chi},\\
&=&\frac{1}{2}\frac{p^{2}_{a}}{m}+\frac{1}{2}\frac{V_{\rm eff}(a)}{M^{2}}p^{2}_{\chi},\label{Hamiltonian newform}
\ea
The  Lift canonical momenta associated with the $\mathcal{H}_{\rm Lift}$
are expressed as:
\ba
{p}_{a}&=&\frac{\partial L_{\rm Lift}}{\partial \dot{a}}=m\dot{a},\\
{p}_{\chi}&=&\frac{\partial L_{\rm Lift}}{\partial \dot{\chi}}=\frac{M^{2}\dot{\chi}}{V_{\rm eff}(a)}=AM=\sqrt{2}M.
\ea
The Hamilton's equation reveals that
\ba
\dot{{p}}_{a}&=&-\frac{\partial \mathcal{H_{\rm Lift}}}{\partial a}=-\frac{p^{2}_{a}}{2M^{2}}V_{\rm eff,a}. \label{form pa}
\ea
Eq.(\ref{form pa}) represents the cosmological acceleration equation as demonstrated in Eq.(\ref{accel eq1}). 
The cyclical nature of $\chi$ implies that its conjugate momentum $p_{\chi}$ is conserved. This allows us to take
\ba
\dot{p}_{\chi}&=&\frac{d p_{\chi}}{dt}=-\frac{\partial \mathcal{H}_{\rm Lift}}{\partial \chi}=0. \label{canonical Pchi}
\ea
Substituting $p_{a},p_{\chi}$ and $V_{\rm eff}$ into Eq.(\ref{Hamiltonian newform}) and imposing the constraint 
$\mathcal{H}_{\rm Lift}=0$, we recover the Friedmann equation as given in Eq.(\ref{Full Nathan Hamiltonian}). This demonstrates that the Eisenhart lift preserves energy conservation in the cosmological system, even after reformulating Rosen's original Lagrangian in the Eisenhart-Lift framework. Employing the methodology for conformal Killing equations, we characterize the complete set of Killing vectors (See Appendix C of \cite{RM Wald}):
\ba
\nabla_{A}\xi_{B}+\nabla_{B}\xi_{A}&=&\frac{2}{n}(\nabla^{C}\xi_{C})G_{AB} ,\label{Killing prove}\\
\nabla_{A}\xi_{B}+\nabla_{B}\xi_{A}&=& \mathcal{F}G_{AB}.
\ea
To obtain the second line of the above equation, we multiply both sides of Eq. (\ref{Killing prove}) by $G^{AB}.$ Due to the symmetry of $G^{AB}$ , this yields:
\ba
2G^{AB}\nabla_{A}\xi_{B}=2\nabla^{B}\xi_{B}.
\ea
We then compare the trace of the left-hand side (LHS) with that of the right-hand side (RHS). Using $G^{AB}G_{AB}=n,$ ,where $n$ is the dimension of the lift metric 
$G^{AB}$ the trace of the RHS gives $n\mathcal{F}$. Equating the traces, we obtain
\ba
2\nabla^{C}\xi_{C}=n\mathcal{F} \quad \Rightarrow \mathcal{F}=\frac{2}{n}\nabla^{C}\xi_{C}.
\ea
In our specific case, we set $n=2$ or the coordinate for the coordinate ${a,\chi}.$ Hence:
\ba
\mathcal{F}=\nabla^{C}\xi_{C}=G^{AB}\nabla_{A}\xi_{B}&=&G^{aa}\nabla_{a}\xi_{a}+G^{\chi\chi}\nabla_{\chi}\xi_{\chi},\\
&=&\frac{1}{m}\partial_{a}\xi_{a}-\frac{1}{2m}\frac{V_{\rm eff,a}}{V_{\rm eff}}\xi_{a}.\label{def F}
\ea
Here, the covariant derivative is defined as:
\ba
\nabla_{A}\xi_{B}=\partial_{A}\xi_{B}-\Gamma^{C}_{AB}\xi_{C}.
\ea
We have three equations associated with the lift Lagrangian, namely:
\ba
2\nabla_{a}\xi_{a}=\mathcal{F} G_{aa}, \label{the first KL}\\
2\nabla_{\chi}\xi_{\chi}=\mathcal{F} G_{\chi\chi}, \label{the second KL}\\
\nabla_{a}\xi_{\chi}+\nabla_{\chi}\xi_{a}=\mathcal{F}G_{a\chi}. \label{the third KL}
\ea
Since $\chi$ does not appear in the metric $G_{AB}$, the Killing vector $\xi_{\chi}$ can be calculated from
\ba
\xi_{(1)}&=&\xi^{\chi}\partial_{\chi}=1\partial_{\chi}=\partial_{\chi},\\
\xi^{\chi}&=&G^{\chi\chi}\xi_{\chi}.
\ea
If we choose $\xi^{\chi}=1$, it can be shown that $\xi_{\chi}=\frac{M^{2}}{V_{\rm eff}(a)}.$ 
The ansatz forms for  the Killing vector component $\xi_{a}$ is 
\ba
\xi_a(a,\chi)=a^{\beta}h(\chi).
\ea
This derivative can be written as
\ba
\partial_a \xi_{a}&=&\beta a^{\beta-1}h(\chi), \quad \quad \partial_{\chi}\xi_{a}=a^{\beta}\frac{\partial h(\chi)}{\partial \chi},\\
\partial_{a}\xi_{\chi}&=&-\frac{M^{2}V_{\rm eff,a}}{V^{2}_{\rm eff}}, \quad \quad  \partial_{\chi}\xi_{\chi}=0.
\ea
Staring from the third conformal Killing equation, this gives
\ba
\partial_{a}\xi_{\chi}+\partial_{\chi}\xi_{a}+\frac{V_{\rm eff,a}M^{2}}{V^{2}_{\rm eff}}=0,\\
\partial_{\chi}\xi_{a}=0  \quad \Rightarrow \xi_{a}(a).
\ea
We know that $\xi_{a}$ depends on $"a"$ only.
From Eq.(\ref{the first KL}), the first Killing equation becomes
\ba
2(\partial_{a}\xi_{a}-\cancel{\Gamma^{c}_{aa}}\xi_{c})&=&\mathcal{F}G_{aa}\nonumber \\
\partial_{a}\xi_{a} &=& -\frac{1}{2}\frac{V_{\rm eff,a}}{V_{\rm eff}}\xi_{a},\\  \Rightarrow \xi_{a}&=&\frac{1}{\sqrt{V_{\rm eff}}}.
\ea
The second conformal Killing equation is 
\ba
2\cancel{\partial_{\chi}\xi_{\chi}}-\frac{M^{2}}{2m}\frac{V_{{\rm eff},a}}{V^{2}_{\rm eff}} \xi_{a}  &=& \frac{M^{2}}{mV_{\rm eff}}\partial_{a}\xi_{a},\nonumber\\
-\frac{V_{{\rm eff},a}}{2V_{\rm eff}}\xi_{a}
&=&\partial_{a}\xi_{a}.
\ea
We then find
\ba
\xi_{a}= \frac{1}{\sqrt{V_{\rm eff}}}.
\ea
This helps confirm the form of $\xi_{a}.$
Substituting $\xi_{a}$ into Eq.(\ref{def F}), one yields the exact form of the conformal factor:
\ba
\mathcal{F}(a)&=&-\frac{1}{m}\frac{V_{\rm eff,a}}{V^{3/2}_{\rm eff}}.
\ea
Keeping just the cosmological constant and curvature terms gives:
\ba
\frac{V_{\rm eff,a}}{V_{\rm eff}}&=&\frac{1+\frac{\lambda}{2}}{\frac{a}{2}+\frac{3k}{2a\Lambda}}=-mF(a)\sqrt{\frac{mk}{2}-\frac{m\Lambda a^{2}}{6}}, \\
F(a)&=&-\frac{2\sqrt{2}}{m^{3/2}}\frac{(1+\frac{\lambda}{2})}{a(1+\frac{3}{a^{2}\Lambda})\sqrt{1-\frac{a^{2}\Lambda}{3}}}.
\ea
Taking the limit $a^{2}\Lambda \ll 1$, this gives
\ba
F_{\Lambda}(a) \approx -\frac{2\sqrt{2}}{3}(1+\frac{\lambda}{2})\Lambda_{0}a^{\lambda+1}.
\ea
For $\Lambda \rm{CDM}$, we have $F_{\Lambda}\, (a) \approx -\frac{2\sqrt{2}}{3}\Lambda_{0}a$. Notably, the conformal factor exhibits a linear dependence of the scale factor upon the cosmological constant.
Next, we use 
to construct the conserved quantity,
\ba
\ell=\xi^{A}p_{A}&=&\xi^{a}p_{a}+\xi^{\chi}p_{\chi},\\
&=&\frac{\dot{a}}{\sqrt{V_{\rm eff}}}+\sqrt{2}M.
\ea
Using the standard equation-of-state relation $\rho(a)=\rho_{0}a^{-3(1+w)},$
for a closed Universe ($k=+1$) with $\rho_{0}=0$, it led to the integral
\ba
\sqrt{\frac{m}{2}}(\ell-\sqrt{2}M)t &=&\int_{0}^{a} \frac{da}{\sqrt{1-\frac{\Lambda_{0}a^{2+\lambda}}{3}}}. \label{const eq}
\ea
For the square root to be real (not imaginary), its argument must satisfy
$1-\frac{\Lambda_{0}}{3}a^{2+\lambda}\geq 0.$ Using the fundamental theorem of calculus together with the chain rule,
\ba
\sqrt{\frac{m}{2}}(\ell-\sqrt{2}M)=\frac{d}{dt}\Bigg(  \int_{0}^{a(t)}\frac{da'}{\sqrt{1-\frac{\Lambda_{0}a'^{2+\lambda}}{3}}}\Bigg)=\frac{\dot{a}}{\sqrt{1-\frac{\Lambda_{0}}{3}a^{2+\lambda}}}\,.
\ea
A recollapse (maximum expansion) means that $\dot{a}=0$ and $\ddot{a}<0$, i.e.,
\ba
\ddot{a} \propto  -\frac{\Lambda_{0}}{6}(2+\lambda)a^{\lambda+1}<0
\ea
where $\Lambda_{0}>0 $ and $\lambda > -2.$
One can readily show that the maximum size of the Universe depends explicitly on $\lambda$, namely,
\ba
a_{\rm max}=\Big(\frac{3}{\Lambda_{0}}\Big)^{\frac{1}{2+\lambda}}.
\ea
If $\Lambda_{0}>0$ and $\lambda<-2$, then $\ddot{a}>0$, indicating that the turning point corresponds to a minimum scale factor rather than a maximum one. In this case, the Universe undergoes a nonsingular bounce, with the minimum size given by
\ba
a_{\rm min}=\Big(\frac{3}{\Lambda_{0}}\Big)^{\frac{1}{2+\lambda}}=\Big( \frac{\Lambda_{0}}{3}\Big)^{\frac{1}{|\lambda|-2}}. 
\ea
The choice of $\lambda$ is therefore of central importance, since it determines both the location and nature of the turning point and whether a bouncing cosmological solution can occur. 
It should be noted that Eq.(\ref{const eq}) can be expressed in terms of the Gaussian hypergeometric function (see p.306 of Ref.\cite{Philippe 1995}) as
\ba
\int \frac{da}{\sqrt{1-c a^{n}}}=a\,\, {}_{2}F_{1}(\frac{1}{2},\frac{1}{n};1+\frac{1}{n};c\,x^{n})
\ea
with $n=2+\lambda$, $ c=\frac{\Lambda_{0}}{3}$.
This gives the exact solution in implicit form
\ba
\sqrt{\frac{m}{2}}(\ell-\sqrt{2}M)\,t=a\,\, {}_{2}F_{1}(\frac{1}{2},\frac{1}{2+\lambda};1+\frac{1}{2+\lambda};\frac{\Lambda_{0}}{3} a^{2+\lambda})
\ea
Interesting, if we set $\lambda=0$ or the cosmological constant, the hypergeometric function reduced to
\ba
{}_{2}F_{1}(\frac{1}{2},\frac{1}{2};\frac{3}{2};c x^{2})=\frac{\sin^{-1} (x)}{x}.
\ea
It leads to 
\ba
a(t)=\sqrt{\frac{3}{\Lambda_{0}}}\sin\Big[ \sqrt{\frac{m\Lambda_{0}}{6}(\ell -\sqrt{2} M \big)}\,\,t \Big].
\ea
Thus, the standard oscillatory closed-Universe solution emerges as a special case of the more general hypergeometric cosmology. In this broader framework, the constant of motion obtained from the Eisenhart lift method plays a dual role: it both rescales the cosmic time. It determines the oscillation period of the universe:
\ba
T=\frac{2\pi}{\sqrt{\frac{m\Lambda_{0}}{6}\big(\ell-\sqrt{2} M \big)}},
\ea
where the maximum Universe in this case is $a_{\rm max}=\sqrt{\frac{3}{\Lambda_{0}}}.$ 
The larger the value of $(\ell-\sqrt{2} M) $, the faster the Universe oscillates, resulting in a shorter period. This quantity emerges naturally from the conserved charges of the Eisenhart lift and encodes the initial expansion rate as well as the effective potential energy of the system.
If we set $m(\ell-\sqrt{2} M )=1$, this gives $T=\frac{2\sqrt{6}\pi}{\sqrt{10^{-35} s^{-2}}} \approx 154 \,\rm {Gyr}. $  
To satisfy the result from Ref{\cite{Neukart 2025}} , which gives the typical full duration of a single expansion–contraction cycle as $T=62.0\pm 2.5\, Gyr,$ we find that $m(\ell-\sqrt{2}M)\approx 15.32$

\section{Eisenhart Lift Approach to the Wheeler-DeWitt Equation}\label{ch5}
\begin{figure}
    \centering
    \includegraphics[width=0.9\linewidth]{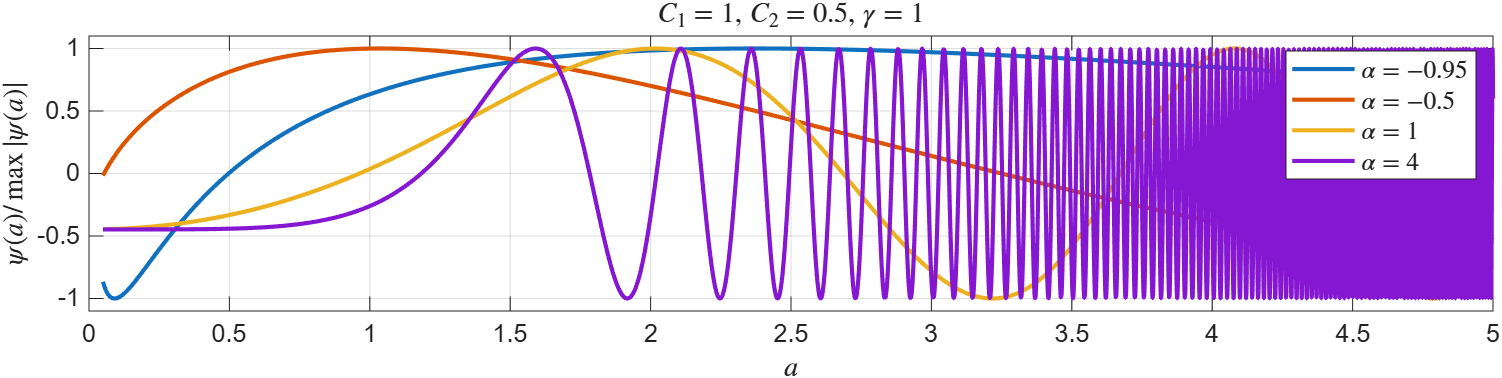}
    \caption{We illustrate the behavior of the solution, Eq.~(\ref{solution}) is plotted for
$C_{1}=1$, $C_{2}=0.5$, and $\gamma=1$, with four different values of the
parameter $\alpha$, namely $\alpha=-0.95$, $-0.5$, $1$, and $4$.}
    \label{fig:placeholder}
\end{figure}

By replacing the conjugate momenta with their operator forms, 
$p_{a}=-i\hbar\frac{\partial}{\partial a}$ and $p_{\chi}=-i\hbar \frac{\partial}{\partial \chi}$ in the lift Hamiltonian, the Wheeler–DeWitt equation becomes:
\ba
\Big[ -\frac{\hbar^{2}}{2m} \frac{\partial^{2}}{\partial a^{2}}-\frac{\hbar^{2}}{2M^{2}}V_{\rm eff} (a)\frac{\partial^{2}}{\partial \chi^{2}}\Big]\Phi(a,\chi)&=&0,\\
{\mathcal{H}}_{\rm Lift}\Psi(a,\chi)&=&0.
\ea
Now we will use the general form of the Wheeler DeWitt equation on minisuperspace
\ba
\Big[ -\frac{\hbar^2}{2}\square+\xi\mathcal{R}  \Big]\Psi(a,\chi)=0.
\ea
One can express the Laplace-Beltrami operator on superspace as
\ba
\square=\frac{1}{\sqrt{G}}\partial_{A}(\sqrt{G} \,G^{AB}\partial_{B})
\ea
Now we transition from the usual pseudo-Riemannian spacetime, where the metric determinant is negative (denoted by $\sqrt{-G}$), to a Riemannian space characterized by a positive-definite metric. In this Euclidean regime, the volume element is given by $\sqrt{G}$, is the determinant of the minisuperspace metric (the DeWitt metric), which the Eisenhart replaces lift metric $G_{AB}$. The Laplace-Beltrami operator in our case is
\ba
\square=\frac{1}{m}\frac{\partial^{2}}{\partial a^{2}}-\frac{{V_{\rm eff,a}}}{2mV_{\rm eff}}\frac{\partial}{\partial a}+\frac{V_{\rm eff}}{M^{2}}\frac{\partial^{2}}{\partial \chi^{2}}.
\ea
The Ricci scalar of configuration space can be derived direcly from $G_{AB}$, this gives
\ba
\mathcal {R}=\frac{2V''_{\rm eff} V_{\rm eff}-3{V'_{\rm eff}}^{2}}{2mV^{2}_{\rm eff}},
\ea
where we use  $V'_{\rm eff}=\frac{\partial V_{\rm eff}}{\partial a}$ and $V''_{\rm eff} =\frac{\partial^{2} V_{\rm eff}}{\partial a^{2}}.$ The Ricci scalar corresponding to an arbitrary spatial curvature parameter $(k)$ and a cosmological constant dominated effective potential can be obtained as
\ba 
\mathcal{R}
=
-\frac{
(\lambda+2)\Lambda_{0}a^{\lambda}
\left[
6k(\lambda+1)
+\Lambda_{0}(\lambda+4)a^{\lambda+2}
\right]
}
{
2m\left(3k-\Lambda_{0}a^{\lambda+2}\right)^{2}
}.
\ea
This reduces to
\ba
\mathcal{R}=-\frac{(\lambda+2)(\lambda+4)}{2ma^{2}}
\ea
when $k=0$, as expected. It is evident that the Ricci scalar diverges in the limit $a\rightarrow 0,$ , behaving as $\mathcal{R}\propto \frac{1}{a^{2}}$. This divergence signals the presence of a curvature singularity in the minisuperspace geometry at a vanishing scale factor. There are a  few special cases: 
\ba
\lambda=-2,\quad \mathcal{R}=0,\\
\lambda=-4, \quad \mathcal{R}=0,\\
-4<\lambda<-2, \quad\mathcal{R}>0,\\
\lambda >-2 \quad \text{and}\, \lambda <-4,\,\, \mathcal{R}<0.
\ea
To be consistent with the Wheeler-DeWitt analysis presented in the following section, we restrict our attention to the parameter region corresponding to a negative minisuperspace Ricci scalar. Using the relation $\alpha=1+\frac{\lambda}{2}$, the condition $\mathcal{R}<0$ is satisfied for $\lambda>-2$, which implies $\alpha>0$, or for $\lambda<-4$, which implies $\alpha<-1$. The limiting cases $\lambda=-2$ and $\lambda=-4$ correspond to a vanishing Ricci scalar, yielding $\alpha=0$ and $\alpha=-1$, respectively. Now we use the choice that \cite{Halliwell_1988}
\ba
\xi=\frac{d-2}{4(d-1)}.
\ea
For the Lift Lagrangian under consideration, the minisuperspace dimension is $d=2$, which implies $\xi = 0$. Consequently, the quantum correction to the potential term proportional to  $\xi \mathcal{R}$ can be neglected. A detailed discussion of this issue can be found on p.249 of \cite{Kiefer}.
The Wheeler-DeWitt equation in Laplace-Beltrami form is
\ba
-\frac{\hbar^{2}}{2m} \Bigg[ \frac{\partial^{2}}{\partial a^{2}}-\frac{V_{\rm eff,a}}{2V_{\rm eff}}\frac{\partial}{\partial a}+\frac{m V_{\rm eff}}{M^{2}}\frac{\partial^{2}}{\partial \chi^{2}} \Bigg]\Psi(a,\chi)=0.
\ea
Since $V_{\text{eff}}$ depends only on the scale factor $a(t)$, we can separate the variables and express the wave function as
\ba
\Psi(a,\chi)=\psi(a)\varphi(\chi)=\psi(a)e^{i \tilde{p} \chi}
\ea
\ba
\Bigg[  \frac{\partial^{2}}{\partial a^{2}}-\frac{V_{\rm eff,a}}{2V_{\rm eff}}\frac{\partial}{\partial a}-\frac{m\tilde{p}^{2} V_{\rm eff}}{M^{2}}\Bigg]\psi(a)&=&0,
\ea
Here, $\tilde{p} =$  the unknown constant. 
In the scenario where the cosmological constant and flat space dominates the universe, 
\ba
\Big(\frac{V'_{\rm eff}}{V_{\rm eff}}\Big)&=&\frac{2}{a}+\frac{\lambda}{a}.\\
\Big[ \frac{\partial^{2}}{\partial a^{2}}-(1+\frac{\lambda}{2} )\frac{1}{a}\frac{\partial }{\partial a}+\gamma  a^{2+\lambda}\Big]\psi(a)&=&0,
\ea
where $\gamma \equiv \frac{m^{2} \tilde{p}^{2} \Lambda_{0}}{6M^{2}}>0$. 
This is a second-order linear ODE for $\psi(a)$, with variable coefficients involving $a.$ To Remove the first-derivative term we define $\alpha=1+\frac{\lambda}{2}$, this yields
\ba\label{WDW1}
\psi''-\frac{\alpha}{a}\psi'+\gamma a^{2\alpha}\psi=0.
\ea
We assume that 
\ba
\psi(a) = u(a)a^{\alpha/2}. 
\ea
The first and second derivatives of this function are given by
\ba
\psi'(a)&=&a^{\alpha/2}u'+\frac{\alpha}{2}a^{\frac{\alpha}{2}-1}u,\\
\psi''(a)&=&a^{\frac{\alpha}{2}}u''+\alpha a^{\frac{\alpha}{2}-1}u'+\frac{\alpha}{2}(\frac{\alpha}{2}-1)a^{\frac{\alpha}{2}-2}u.
\ea
Substituting them into Eq.(\ref{WDW1}) and canceling the term contribution to first derivative, i.e.
\ba
a^{\frac{\alpha}{2}}u''+\cancel{\frac{\alpha}{2}a^{\frac{\alpha}{2}-1}u' }+\frac{\alpha}{2}(\frac{\alpha}{2}-1)a^{\frac{\alpha}{2}-2}u-\cancel{\frac{\alpha}{2}a^{\frac{\alpha}{2}-1}u' }-\frac{\alpha^{2}}{2a}a^{\frac{\alpha}{2}-1}u+\gamma a^{2+\lambda}u(a)a^{\frac{\alpha}{2}}&=&0.\label{eqw}
\ea
We can rearrange Eq.(\ref{eqw}) into the simple form shown below
\ba\label{WDW2}
u''(a)+\Big(\gamma  a^{2\alpha}-\frac{\alpha(\alpha+2)}{4a^{2}}\Big)u(a)&=&0.
\ea
Introducing the new variable $x=\frac{\sqrt{\gamma}}{\alpha+1}a^{\alpha+1},\,\,\alpha \neq -1,$ or $\lambda \neq -4 $, we come up with
\ba
\frac{dx}{da}&=&\sqrt{\gamma}a^{\alpha},\\
\frac{du}{da}&=&\frac{du}{dx}\frac{dx}{da},\\
\frac{d}{da}&=&\sqrt{\gamma}a^{\alpha}\frac{d}{dx}.
\ea
and then
\ba
u_{xx}+\frac{\alpha}{\sqrt{\gamma}}a^{-(\alpha+1)}u_{x}+\Bigg[ 1-\frac{\beta}{\gamma } \Bigg]u&=&0.
\ea
Choosing $\beta=\frac{\alpha(\alpha+2)}{4(\alpha+1)^{2}}$ and $u(x)=x^{m}y(x)$ so that the equation becomes standard Bessel form. This gives
\ba
y''+\frac{(2m+\frac{\alpha}{\alpha+1})}{x}y'+\Big[ 1+\frac{(m(m-1)+\frac{\alpha m}{(\alpha+1)})-\beta}{x^{2}} \Big]y=0.
\ea
Matching the Bessel equation, see p.322 of Ref\cite{Philippe 1995}, we find
\ba
y''+\frac{1}{x}y'+(1-\frac{\nu^{2}}{x^{2}})y=0\,,
\ea
whose solutions take the form
\ba
y(x)=C_{1}J_{\nu}(x)+C_{2}Y_{\nu}(x)\,,
\ea
where $J_{\nu}$ is a Bessel function of the first kind and $Y_{\nu} (x)$ is a Bessel function of the second kind, $C_{1}$ and $C_{2}$ are constant.
By allowing 
\ba
\frac{1}{x}=\frac{2m+\frac{\alpha}{(\alpha+1)}}{x} \rightarrow m=\frac{1}{2(\alpha+1)}\quad \alpha \neq -1 \,,
\ea
and matching the Bessel index for the coefficient of $\frac{1}{x^{2}}$
 \ba
 \frac{\nu^{2}}{x^{2}}=-\frac{(m(m-1)+\frac{\alpha m}{(\alpha+1)})-\beta}{x^{2}},
 \ea
 it is trivial to show that
 \ba
 {(m(m-1)+\frac{\alpha m}{(\alpha+1)})-\beta}=-\frac{(2\alpha+1)}{4(\alpha+1)^{2}}+\frac{2\alpha}{4(\alpha+1)^{2}}-\frac{\alpha(\alpha+2)}{4(\alpha+1)^{2}}=-\frac{1}{4}.
 \ea
 Therefore, it can be concluded that $\nu^{2}=\frac{1}{4},$ and hence $\nu=\frac{1}{2}.$ Thus, in the half-integral order case, the Bessel functions reduce to elementary trigonometric functions:
 \ba
 y(x)=C_{1}J_{1/2}(x) +C_{2}Y_{1/2}(x)
 \ea
 where 
 \ba
 J_{1/2}(x)=\sqrt{\frac{2}{\pi x}}\sin x,\\
 Y_{1/2}(x)=-\sqrt{\frac{2}{\pi x}} \cos x.
 \ea
 This gives
 \ba
\psi(a)
=
a^{\frac{\alpha}{2}}
\Big(
\frac{\sqrt{\gamma}}{\alpha+1}a^{\alpha+1}
\Big)^{\frac{1}{2(\alpha+1)}-\frac{1}{2}}
\sqrt{\frac{2}{\pi}}
\Bigg[
C_{1}\sin\!\Big(
\frac{\sqrt{\gamma}}{\alpha+1}a^{\alpha+1}
\Big)
-
C_{2}\cos\!\Big(
\frac{\sqrt{\gamma}}{\alpha+1}a^{\alpha+1}
\Big)
\Bigg]\,.
\label{solution}
\ea
 From Fig.\ref{fig:placeholder}, it is evident that increasing $\alpha$ or $\lambda$ leads to more rapid oscillations of the Wheeler–DeWitt wavefunction at large values of the scale factor $a$, indicating a stronger semiclassical behavior and a closer correspondence with a classical cosmological evolution. In contrast, negative values of $\alpha$ enhance the prominence of quantum effects in the small-$a$ regime, where the universe is expected to be dominated by quantum gravitational dynamics.
\section{Concluding Remarks}\label{ch6}
Rosen's Lagrangian serves as the central framework of the present analysis and provides valuable insights into the dynamics and large-scale evolution of the Universe. As demonstrated in this work, a cosmological model originally described by a single dynamical degree of freedom can be naturally extended to a higher-dimensional configuration space through the introduction of an auxiliary field $\chi$, in the spirit of the Eisenhart lift formalism. This construction complements the conventional treatment involving scalar fields and enables a geometrical formulation of the cosmological dynamics. Within this framework, the conformal Killing equations determine the conformal factor as $\mathcal{F}(a)=-\frac{1}{m}\tfrac{V_{{\rm eff},a}}{V_{\rm eff}^{3/2}}$, establishing a direct connection between the spacetime symmetry and the effective potential governing the evolution. Furthermore, the maximum size reached by the Universe is found to depend explicitly on the parameter $\lambda$, yielding $a_{\rm max}=\left(\frac{3}{\Lambda_0}\right)^{1/(2+\lambda)}$. The resulting cosmological evolution exhibits a cyclic behavior characterized by alternating phases of expansion and contraction, with a maximum cosmic period of approximately $154,{\rm Gyr}$. Finally, by constructing the Wheeler-DeWitt metric through the auxiliary field $\chi$, we obtain analytical quantum cosmological solutions that remain physically relevant in both the early inflationary epoch and the late-time accelerated era of the Universe. In particular, acceptable quantum states arise for $\alpha>0$, corresponding to the condition $\lambda>-2$, thereby identifying the parameter region in which the model remains both mathematically consistent and physically viable. 

Another promising direction is to investigate the interplay between Eisenhart-lift symmetries and Noether symmetries in Rosen cosmology. Since both approaches are closely related to the geometric structure of the configuration space, a combined analysis may provide a systematic method for identifying integrable cosmological models and constructing exact quantum solutions. Moreover, the extension of the present framework to quantum cosmology with non-minimally coupled scalar fields or modified theories of gravity may offer new insights into the emergence of classical spacetime from the underlying Wheeler-DeWitt dynamics, see e.g., \cite{Alberghi:1999gh,Singh:2025spz}.

\acknowledgments
Thailand's NSRF financially supports the work of P.C. via PMU-B under grant number PCB37G6600138. The National Research Council of Thailand (NRCT) funds S.M. and N.K. under contract number N42A660971.


\begin{thebibliography}{99}

\bibitem{Eisenhart origin}
L. P. Eisenhart, Dynamical trajectories and geodesics, Ann Math \textbf{30} (1/4): 591–606 (1928).

\bibitem{Rosen1993}
N. Rosen, (1993). Quantum   Mechanics of a MiniUniverse. International Journal of Theoretical Physics, Vol. 32, No. 8, 1993.

\bibitem{Newtonian cosmo}
H. S. Vieira and V. B. Bezerra. 2014 Lagrangian formulation of Newtonian cosmology. Rev. Bras. Ensino Fís. 36 (3) 

\bibitem{Adiabatic invariant}
N. Kaewkhao and P. Channuie,(2023). Adiabatic invariant approach on Fridmann cyclic Universe. Nuclear Physics.\textbf{B} 987 116088.

\bibitem{Cariglia 2018} 
M. Cariglia, A. Galajinsky, G.W. Gibbons, and P.A. Horvathy. Cosmological aspects of the Eisenhart-Duval lift. Eur.Phys.J.C.,78(4): 314,2018.


\bibitem{Georgios 2023} 
G. K. Karananas,M. Michel,and J. Rubio.The geometry of inflationary observables: Lifts, flows, equivalence classes. Phys.Lett.B 850 (2024) 138524

\bibitem{Macedo 2023}
H.A. P. Macedo, L.S. Brito, J.F. Jesus, and M.E. S. Alves. Cosmological constraints on $\Lambda(t)$ CDM models. Eur. Phys. J.C  (2023) 83:1144. 

\bibitem{Koussour 2024}
M. Koussour, N. Myrzakulov, and Javlon Rayimbaev.Cosmological constraints on time-varying cosmological terms: A study of FLRW Universe models with $\Lambda(t)$ CDM cosmology  Adv.Space Res. 74 (2024) 1343-1351

\bibitem{Overduin 1998}
J.M. Overduin, F.I. Cooperstock, Evolution of the scale factor with a variable cosmological term. Phys.Rev.D 58 (1998) 043506

\bibitem{Carneiro 2021}
S. Carneiro, M.A. Dantas, C. Pigozzo, J.S. Alcaniz. Observational constraints on late-time Lambda(t) cosmology.Phys.Rev.D 77 (2008) 083504.

\bibitem{Saibal Ray}
S. Ray, F. Rahaman, U. Mukhopadhyay and R. Sarkar. Variable Equation of State for Generalised Dark Energy Model
Int.J.Theor.Phys. \textbf{50} (2011) 2687-2696.


\bibitem{Wei Chen 1990}
W. Chen and Y.S. Wu. Implication of a cosmological constant vary as $R^2$, Phys. Rev. D 41, 695 (1990), Erratum Phys. Rev. D 45, 4728 (1992)(1990).

\bibitem{Myrzakulov 2024}
Y. Myrzakulov, M. Koussour, M. Karimov, and J. Rayimbaev.
Signature flips in time-varying $\Lambda(t)$
 cosmological models with observational data. Eur.Phys.J.C 84 (2024) 7, 665
Published: Jul 5, 2024 

\bibitem{A Liddle 2015}
A. Liddle. An Introduction to Modern Cosmology,Edition, 3 ; Publisher, John Wiley \& Sons, 2015.

\bibitem{Finn 2023 Book}
K. Finn, Geometric Approaches to Quantum Field Theory, Springer , Switzerland (2021).

\bibitem{Cariglia 2015}
M. Cariglia,  F.K.  Alves, The Eisenhart lift: a didactical introduction of modern geometrical concepts from Hamiltonian dynamics, Eur J Phys \textbf{36} (2):025018 (2015). 


\bibitem{K Fin 2018}
K. Finn, S. Karamitsos  and A. Pilaftsis,Eisenhart lift for field theories, Phys. Rev. D \textbf{98}, 1 016015 (2018).

 \bibitem{Finn 2019}
K. Finn, S. Karamitsos, Finite measure for the initial conditions of inflation, Phys. Rev. D, \textbf{99}, 063515. (2019).


\bibitem{RM Wald}
R. M. Wald. General Relativity. The University of Chicago Press. (1984)




\bibitem{Philippe 1995}
P.  Dennery and A.  Krzywicki, Mathematics for Physicists. Dover Publications, INC. New York (1995).


\bibitem{Neukart 2025}
F. Neukart, E. Mark, V, Vinokur. Counting Cosmic Cycles: Past Big
Crunches, Future Recurrence Limits,
and the Age of the Quantum Memory
Matrix Universe. Entropy 27(10), 1043, (2025).

\bibitem{Halliwell_1988}
J.J. Halliwell. Derivation of the Wheeler-DeWitt equation from a path integral for minisuperspace models. Phys. Rev. D\textbf{38}, 8 (1988).

\bibitem{Kiefer}
C. Kiefer, Quantum Gravity. $2^{nd}$ Edition. Oxford University Press. (2007).

\bibitem{Alberghi:1999gh}
G.~L.~Alberghi, R.~Casadio and A.~Gruppuso,
Phys. Rev. D \textbf{61} (2000), 084009

\bibitem{Singh:2025spz}
H.~Singh and M.~K.~Nandy,
Nucl. Phys. B \textbf{1022} (2026), 117219

\end{thebibliography}
\end{document}